# Strain-induced modification of spin-optical dynamics in silicon vacancy centers for integrated quantum technologies

Maximilian Hollendonner[1], Fedor Dzmitryevich Hrunski[1], Daniel Scheller[1], Kim Ullerich[1], Shravan Kumar Parthasarathy[1,2], Wolfgang Knolle[3], Maximilian Schober[4], Mirjam Neubauer[4], Durga Bhaktavatsala Rao Dasari[5], Michel Bockstedte[4], and Roland Nagy[1]

[1]Institute of Applied Quantum Technologies (AQuT.), Friedrich-Alexander University Erlangen-Nürnberg, Erlangen, Germany

[2]Fraunhofer Institute for Integrated Systems and Device Technology (IISB), Erlangen, Germany

[3]Department of Sensoric Surfaces and Functional Interfaces, Leibniz-Institute of Surface Engineering (IOM), Leipzig, Germany

[4]Institute for Theoretical Physics, Johannes Kepler University Linz, Linz, Austria

[5]3rd Institute of Physics, University of Stuttgart, ZAQuant, Stuttgart, Germany

Corresponding author: Roland Nagy, roland.nagy@fau.de

**Abstract**

Silicon vacancy ($V_{Si}$) centers in 4H silicon carbide have emerged as a highly promising platform for semiconductor-based quantum technologies, combining excellent spin and optical properties with an industrial-grade, CMOS-compatible material. As these defects are increasingly integrated into practical quantum devices, they inevitably encounter lattice strain. However, while the impact of strain is well documented for other solid-state defects like NV centers in diamond, its specific influence on key $V_{Si}$ spin dynamics such as initialization fidelity and state lifetimes remain largely unexplored. In this work, we address this critical gap by designing fully optical pulse sequences and incorporating the effective spin-3/2 strain Hamiltonian into our analysis. This combined approach allows us to isolate both axial and transverse strain contributions and systematically characterize their effect on the metastable state transition rates. Specifically, we reveal that strain significantly reduces the transition rates from the energetically lowest metastable state to the ground state quartet, leading to decreased photon emission. Supported by first-principles calculations, our findings provide a deeper understanding of $V_{Si}$ spin-strain dynamics, yielding crucial insights for the robust deployment of these centers in realistic, strain-prone environments.

**Introduction**

A central requirement for scalable quantum technologies is the realization of efficient and coherent spin–photon interfaces. Achieving this goal relies on progress along two complementary directions. On the one hand, advances in material engineering—such as isotopic purification [1], controlled defect creation [2,3] and electric tuning of the resonances [4] – enable the formation of multiple optically active spin centers with long-lived nuclear spin registers [5–9] in their vicinity. On the other hand, the performance of these systems is critically determined by their optical properties, including transition linewidths, metastable state dynamics, and branching ratios [10,11]. These factors directly govern spin initialization and readout fidelities [12,13], and therefore set fundamental limits on the achievable performance of quantum protocols.

In practice, however, efficient light–matter interaction remains a major bottleneck. The inherently low photon collection efficiency of solid-state emitters necessitates their integration into photonic structures [14,15], such as waveguides [16,17] and resonators [3,18,19], where the Purcell effect can enhance emission into the zero-phonon line and improve photon extraction. This is particularly important for relatively dim emitters such as the silicon vacancy centers in 4H-SiC (in the following denoted as $V_{Si}$), for which photonic integration becomes essential rather than optional. However, embedding these emitters into nanophotonic environments [2,16,17] introduces new challenges. Fabrication processes and geometric confinement can induce local lattice strain [17], which modify the electronic structure of the defect.

Despite significant progress in both material development and photonic integration and an enhanced understanding of strain-induced spectral shifts [20], a comprehensive understanding of strain-induced modifications to the $V_{Si}$ spin–optical dynamics remains incomplete. Seminal experiments by [10] allowed for the characterization of the metastable state dynamics of unstrained $V_{Si}$ centers, revealing two effective metastable states between the ground and first excited state quartet. In the present study, we build upon these results by designing fully optical pulse sequences that allow for the characterization of these aforementioned $V_{Si}$ properties under strain. By combining these experimental techniques with an analysis based on an effective spin-3/2 strain Hamiltonian, we are able to characterize both axial and transverse strain contributions and their impact on the metastable state transition rates. Specifically, we find that the transition rates from MS$_1$ [Fig. 1(a)] to the ground state quartet are reduced. Our findings are supported by first-principles calculations and allow for a deeper understanding of $V_{Si}$ spin dynamics and photon emission properties in environments where strain is inevitable, such as in waveguides with tapered fibers.

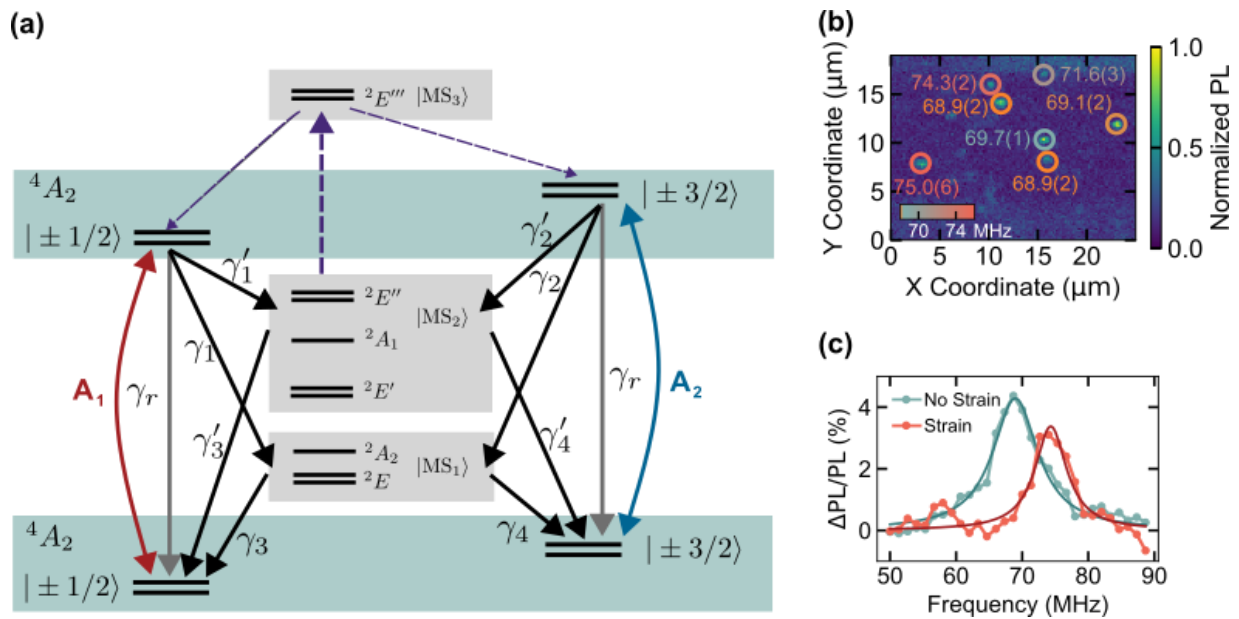

*Figure 1*: (a) Effective 10-level model of the $V_{Si}$ with three effective metastable states. Ground and excited states are connected by two optical transitions, denoted as $A_1$ and $A_2$. From $^4A_2$ excited state, radiative decay happens at rate $\gamma_r$ (gray) to $^4A_2$ ground state. Intersystem-crossing (ISC) rates in and out of the metastable states (black) are not accompanied by photon emission. Laser excitation of the $V_{Si}$ center is accompanied by power-dependent deshelving from $|MS_2\rangle$ to $|MS_3\rangle$ (purple). The indicated doublet states result from our ab initio theory (cf. SI2). (b) Exemplary confocal scan of multiple color centers and their zero-field splitting values (in MHz), where deviations from 70 MHz reveal the presence of strain. (c) Off-resonant ODMR measurements as a function of the applied microwave frequency. Deviations from 70 MHz, reveal whether the color center is subject to strain.

**Experimental investigation of spin dynamics under strain**

Our measurements were performed in a home-built confocal microscope with a cryostat operated at 4 K. For our measurements a bulk c-plane sample is used where a relatively strong variation in the optically detected magnetic resonance (ODMR) peaks is observed, indicating unstrained (peak at 70MHz) and strained (deviations from 70MHz) color centers [Fig. 1(b)]. We use a laser at 730 nm for off-resonant excitation and a tunable diode laser at 916 nm for resonant excitation along the $A_{1/2}$ transitions between ground and first excited $^4A_2$ quartets [Fig. 1(a)]. Interleaved driving of the $A_{1/2}$ transitions was achieved via the sidebands of a phase modulating electro-optic modulator (EOM). The emitted phonon sideband was detected with a superconducting nanowire single photon detector (SNSPD). Further details about the experimental setup can be found in the Supplementary Information (SI1).

The silicon vacancy center possesses spin quartets in ground and first excited state and spin doublet metastable states [Fig. 1(a)] [21–24]. In absence of external magnetic, electric or strain fields, the degenerate $\pm 1/2$ and $\pm 3/2$ spin states of ground (g) and first excited (e) state are split by the zero-field splitting $D_g = 70$ MHz and $D_e \approx 1$ GHz, respectively. From our quantum embedding defect theory (CI-cRPA), we obtain intermediate metastable spin doublets, consisting of three groups of (closely) degenerate states with energy positions that are schematically indicated and discussed in detail in SI2. We also find groups of quartet and doublet hole resonances with strong radiative coupling to the excited $^4A_2$ quartet and $^2E'$ doublet. Moreover, we find that $\gamma_{1,2}$ and $\gamma'_{3,4}$ represent effective (higher-order) rates, which involve a combination of spin-orbit coupling, radiative and/or non-radiative processes and are beyond current modelling capabilities. Based on previous studies [10], we resort to the full $V_{Si}$ spin dynamics under strain being governed by ground and excited state Hamiltonian, the radiative decay rate $\gamma_r$, and the intersystem-crossing (ISC) rates [see Fig. 1(a)] and discuss the equivalence between both descriptions in SI2. The multiplet energies are given in Tab. (SI) 1.

As the complete strain tensor at the defect site is generally difficult to determine, we restrict our analysis to an effective spin–strain Hamiltonian, where the relevant coupling parameters are treated as phenomenological quantities to be extracted from the experiment. Building on established symmetry arguments for spin–strain coupling in $C_{3v}$ symmetry systems, we derive the strain Hamiltonian for the $V_{Si}$ center in 4H-SiC (see SI3). While the general structure follows known results [25,26], an explicit formulation for this defect is, to the best of our knowledge, not available in the literature. The resulting strain Hamiltonian takes the familiar form:

$$H_{g,e} = (D_{g,e} + \Pi_z) \cdot S_z^2 + \left[ \frac{\Pi^{(1)}}{2}(S_+ S_z + S_z S_+) + h.c. \right] + \left[ \frac{\Pi^{(2)}}{2} e^{-i\theta} S_+^2 + h.c. \right]. \quad (1)$$

Here $S_{x,y,z}$ are spin-3/2 operators with $S_\pm = S_x \pm i S_y$. The axial strain component $\Pi_z$ modifies the effective zero-field splitting, while the transverse strain components $\Pi^{(1,2)} \geq 0$ induce spin transitions with Δms=±1 and Δms=±2, respectively. The phase θ∈[0,π) describes the orientation of the transverse strain in the plane perpendicular to the symmetry axis.

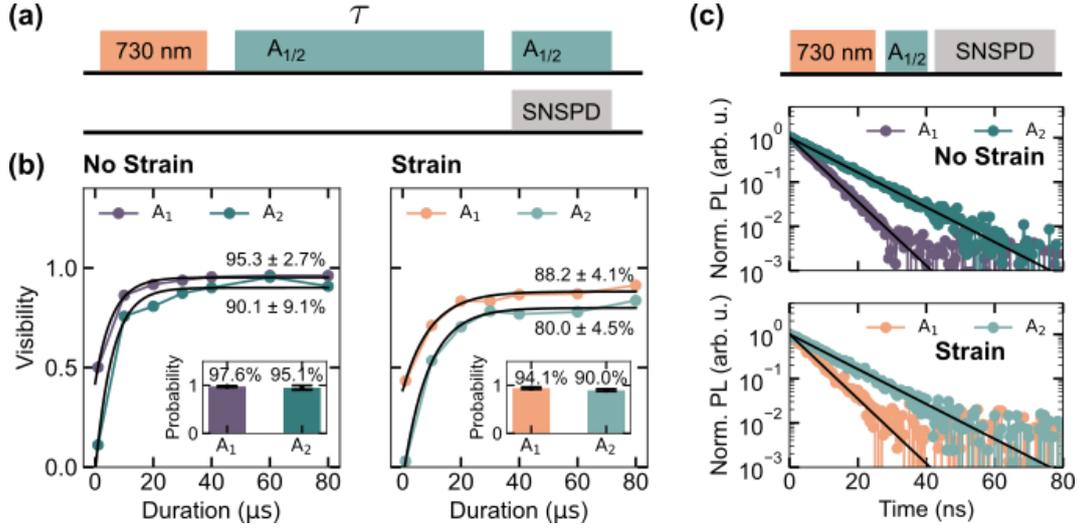

*Figure 2:* (a) Pulse sequence for ground state visibility measurements. The ground state population is inferred purely optically by an initial 730 nm laser pulse to ensure the correct charge state. The visibility after resonant illumination along $A_{1/2}$ for duration $\tau$ is inferred by probing the $m_s = \pm\frac{1}{2}, \pm\frac{3}{2}$ populations with a second laser pulse. Comparison of the detected photoluminescence yields the visibility which allows to deduce the ground state initialization fidelities (insets). (b) Obtained visibilities with and without strain. Insets are the ground state probabilities for the visibilities as written in the main figure. (c) Excited state lifetime measurements. After charge state initialization and spin state mixing, a 1ns sub-lifetime resonant laser pulse excites the $V_{Si}$ center. Measurements of the decaying photoluminescence allow to infer the excited state lifetimes of $|e, \pm 1/2\rangle$ ($A_1$) and $|e, \pm 3/2\rangle$ ($A_2$).

Diagonalization of Eq. (1) shows that the strain coupling parameters lead to deviations of the effective ground state zero-field splitting. Even though this can be measured by ODMR [Fig. 1(c)], the position of the ODMR peak alone does not allow to distinguish between purely axial ($\Pi_z$) and contributions from transverse ($\Pi^{(1,2)}$) strain. Nevertheless, it can be used that transverse strain causes hybridization of the pure ground state eigenstates, an effect which leads to reduced ground state visibilities [12]. Contrary to previous approaches [10,27], we employ a purely optical readout scheme to measure the ground state visibility under resonant excitation. As shown in Fig. 2(a), the pulse sequence consists of an initial laser pulse at 730 nm which ensures the negative charge state. Afterwards the $V_{Si}$ is excited along $A_1$ or $A_2$ for up to 80 μs. The ground state visibility, defined as

$$V = \frac{\left| p^g_{\pm\frac{3}{2}} - p^g_{\pm\frac{1}{2}} \right|}{p^g_{\pm\frac{3}{2}} + p^g_{\pm\frac{1}{2}}} \quad (2)$$

can be obtained by probing the ground state probabilities $p^g_{m_s}$ ($m_s = \pm\frac{1}{2}, \pm\frac{3}{2}$) with short (300 ns) laser pulses along, respectively, $A_1$ ($p^g_{\pm\frac{1}{2}}$) or $A_2$ ($p^g_{\pm\frac{3}{2}}$). With this scheme, we find visibilities of $95.3 \pm 2.7$ % for initialization with $A_1$ and $90.1 \pm 9.1$ % if the $V_{Si}$ is driven along $A_2$ in the absence of strain. Eq. (2) can be used to obtain the ground state probabilities, yielding $p_{\pm 3/2} = 97.6 \pm 1.3$ % ($A_1$) and $p_{\pm 1/2} = 95.1 \pm 4.5$ % ($A_2$), as shown in the insets of Fig. 2(b). These values are in good agreement with previous studies [21,24]. As discussed in the supplementary material (SI3), these fidelities are mainly limited by the dark count rate of the SNSPD and a ground state visibility of $98.04 \pm 2.53$ % is possible if the measured signal is corrected for 7 Hz dark counts, representing a realistic estimate of our SNSPD. The analysis of strained $V_{Si}$ reveals that initialization fidelities are lowered towards $p_{\pm 3/2} = 94.1 \pm 2.1$ % ($A_1$) and $p_{\pm 1/2} = 90.0 \pm 2.3$ % ($A_2$) due to mixing of the ground state spin states. To obtain the strain parameters from Eq. (1), we simultaneously fit the ODMR peak position and the steady-state ground state spin fidelities. As can be seen from Tab. 1, we find that a $V_{Si}$ showing a shifted ODMR as depicted in Fig. 1(c) can have axial and transverse strain contributions.

| $\Pi_z$ (MHz) | $\Pi^{(1)}$ (MHz) | $\Pi^{(2)}$ (MHz) | $\theta$ (rad.) |
|---|---|---|---|
| 1.51 | 3.78 | 3.68 | $0.92\pi$ |

*Table 1: Spin-Strain coupling parameters from fitting Eq. (1) on ODMR and ground state spin visibility measurements.*

Having characterized the contribution of strain as it enters into the spin-strain Hamiltonian [Eq. (1)], we would like to investigate the effect on the radiative and non-radiative transition rates between ground and first excited state manifolds, as well as the effects on the metastable state dynamics. For this reason, a series of purely optical pulse-probe measurements were performed, which allow for the indirect investigation of the ISC rate through fits of the Lindblad master equation (see below).

The rates from the excited states into the metastable states ($\gamma_{1,2}$ and $\gamma'_{1,2}$) in combination with the radiative decay rate $\gamma_r$ [see Fig. 1(a)] determine the excited state lifetimes of, respectively, $|e, \pm 1/2\rangle$ and $|e, \pm 3/2\rangle$ as $\tau^e_{\pm 1/2} = [\gamma_r + \gamma_1 + \gamma'_1]^{-1}$ and $\tau^e_{\pm 3/2} = [\gamma_r + \gamma_2 + \gamma'_2]^{-1}$ [28]. Their value can be measured by initializing the V$_{Si}$ with an initial 730 nm laser pulse, followed by 1 ns resonant laser pulses along A$_1$ (for lifetimes of $|e, \pm 1/2\rangle$) or A$_2$ ($|e, \pm 3/2\rangle$) [Fig. 2(c)]. The emitted photoluminescence decays exponentially with $\tau^e_{\pm 1/2} = 6.09 \pm 0.06$ ns in both cases and $\tau^e_{\pm 3/2} = 11.13 \pm 0.14$ ns ($\tau^e_{\pm 3/2} = 10.43 \pm 0.12$ ns) in presence (absence) of strain. Both values agree well with previous studies [10] and indicate that rates from the excited state quartet into the metastable states are not altered by strain.

**Measurements of metastable state dynamics**

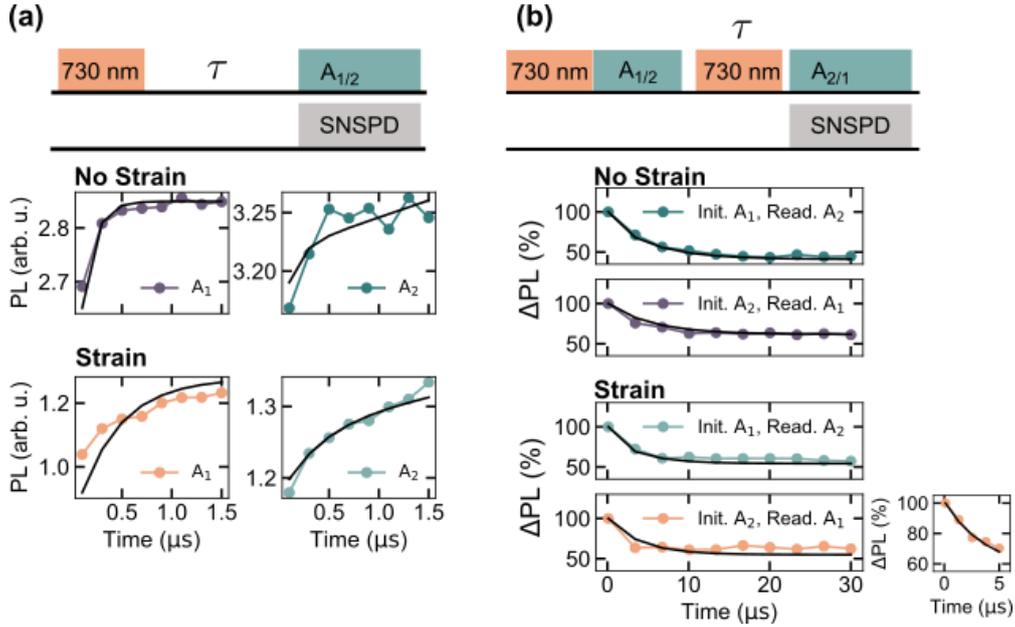

*Figure 3: (a) An initial 730 nm laser pulse (815 μW) pumps the system into the metastable states. The decay into $|g, \pm 1/2\rangle$ and $|g, \pm 3/2\rangle$ manifolds can be measured after delayed readout, resonant with A$_{1/2}$. As saturation is reached slower in presence of strain, the decay rates from the metastable states to the ground states are lowered. (b) The ground state population after off-resonant illumination at 730 nm (50 μW) can be measured by first initializing the system into either $|g, \pm 1/2\rangle$ by the laser resonant with A$_2$ or $|g, \pm 3/2\rangle$ with A$_1$. The ground state populations are then probed with the opposite resonance after having applied the 730 nm laser for variable durations from 100 ns to 30 μs. The displayed change in photoluminescence with respect to the first data point (ΔPL) reveals the steady-state ground state polarizations. Inset: Repeated measurement of initialization with A$_2$ and readout with A$_1$ for shorter durations with which the system is pumped off-resonantly.*

To study the effect of strain on the metastable state dynamics, we extend the previous approach of microwave-assisted readout of ground state spin contrasts [10] towards a fully optical pulse-probe measurement scheme [Fig. 3]. The decay from the metastable states into the ground states, mediated by $\gamma_{3,4}$ and $\gamma'_{3,4}$ [Fig. 1(a)], can be probed by pumping the V$_{Si}$ into both metastable states with an initial

730 nm laser pulse and probing the ground state populations of $|g, \pm 1/2\rangle$ and $|g, \pm 3/2\rangle$ with short (300 ns) laser pulses along $A_{1/2}$ [Fig. 3(a)]. This decay happens at much lower time scales in presence of strain [Fig. 3(a), lower panel], compared to the case when the $V_{Si}$ is not subject to strain [Fig. 3(a), upper panel]. In particular, the total metastable state lifetime of 247 ns, being in accordance with recent findings [24], is increased to 833 ns due to strain.

Besides their total lifetime, the transition rates from metastable to ground states also crucially determine the ground state spin polarization. As previously shown, a steady-state spin polarization of 0.57/0.43 between $|g, \pm 1/2\rangle$ and $|g, \pm 3/2\rangle$ is usually reached upon off-resonant 730nm illumination of the $V_{Si}$ under microwave drive at 70 MHz [10], giving rise to ODMR contrasts of $\approx 3 - 4$ % [see e.g. Fig. 1(c)]. To measure the ground state polarization in presence of strain as well as the time scale after which it is reached, we designed a purely optical pulse sequence as depicted in Fig. 3(b). First, the system is initialized into either $|g, \pm 1/2\rangle$ or $|g, \pm 3/2\rangle$. This is achieved by an initial 730 nm laser pulse which ensures the correct charge state, followed by illumination with EOM laser sidebands either at $A_1$ or $A_2$. Afterwards, the duration with which the $V_{Si}$ is driven by the off-resonant laser is varied from 100 ns up to 30 µs before the $|g, \pm 1/2\rangle$ and $|g, \pm 3/2\rangle$ manifolds are read out spin-selectively through resonant excitation with the opposite transition as the one used for spin initialization. One can see from Fig. 3(b) that after 10-15 µs a steady-state of the detected photoluminescence is reached. The relative change in photoluminescence ($\Delta PL$) with respect to the first data point reveals ground state polarizations of 0.60/0.40 of, respectively, $|g, \pm 1/2\rangle$ compared to $|g, \pm 3/2\rangle$ in absence of strain and 0.53/0.47 in presence of strain (see SI 5 and Tab. (SI) 4).

**Power-dependent deshelving to energetically higher metastable states**

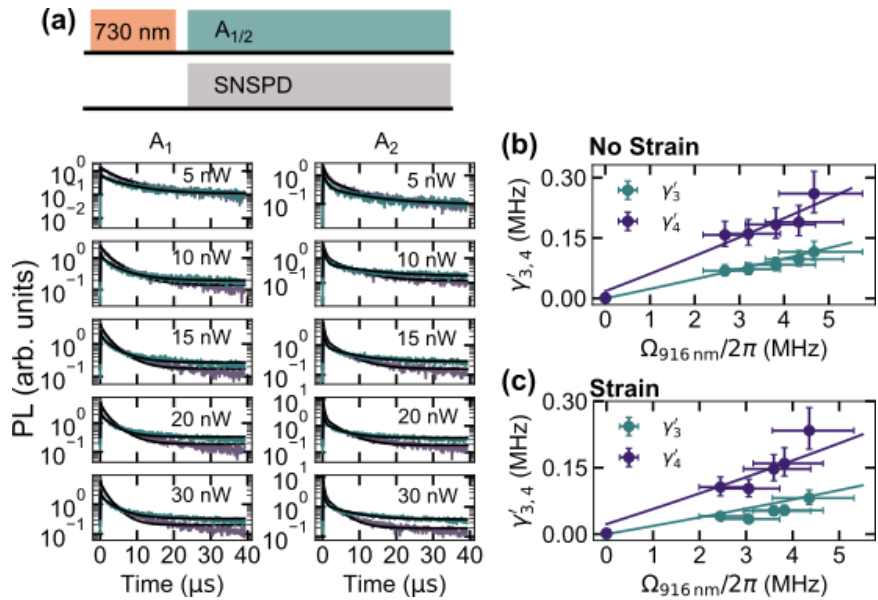

*Figure 4:* (a) Resonant spin depletion. First, the system is initialized into a ground state steady-state through the 730 nm (50 µW) laser, applied for 40 µs. After a short delay of 2 µs, the $V_{Si}$ is pumped into $|g, \pm 3/2\rangle$ ($|g, \pm 1/2\rangle$) by illuminating with $A_1$ ($A_2$) for 40 µs at powers from 5 nW to 30 nW. Purple (green) are measurements without (with) strain. Through fitting the Lindblad master equation on our data, we derive the power-dependent deshelving rates from $MS_2$ to the ground state manifold of (b) no strain and (c) the strain as present within this study. The data points at $\Omega_{916nm} = 0$ were extracted from the measurements in Fig. 3(a) where during the free evolution interval $\tau$ all lasers are switched off. Solid lines in (b) and (c) are fits of Eq. (3).

Previous studies [10] of the $V_{Si}$ spin-dynamics revealed power-dependent deshelving from $MS_2$ to $MS_3$, which manifests itself as power-dependent rates from $|MS_2\rangle$ to the $|^3A_2\rangle$ ground state [Fig. 1(a)]. To study if strain affects this mechanism, we probe the resonant spin depletion by first initializing the $V_{Si}$ into the 0.60/0.40 (0.53/0.47) ground state population with an initial off-resonant laser pulse at 50 µW [Fig. 4(a)]. A short waiting time of 2 µs ensures that the population has decayed from metastable to ground states. While the $V_{Si}$ is illuminated with laser light resonant with either A$_1$ or A$_2$, we measure the emitted photoluminescence of strained and unstrained color centers. The signal decays due to pumping into the opposite spin manifold, i.e. to $|g, \pm 3/2\rangle$ if measurements are performed with A$_1$ and $|g, \pm 1/2\rangle$ for A$_2$. To measure the power-dependent deshelving from MS$_2$ to MS$_3$, the power of the 916 nm laser is varied from 5 nW to 30 nW. From fits of the Lindblad master equation (see below), we do not find substantial strain-induced differences. It can be seen from Fig. 4(b,c) that $\gamma'_{3,4}$ increases linearly with increasing 916 nm Rabi frequencies. From the measurements probing the decay from metastable to ground states [Fig. 3(a)] we can additionally determine the deshelving if all lasers are switched off. With this we find that $\gamma'_{3,4}(\Omega_{916\,nm} = 0) \approx 0$.

To study this power dependence further, we derive a laser-power dependent expression for the deshelving through adiabatic elimination (SI 9) and find

$$\gamma'_{34}(\Omega) \approx \gamma'^{(0)}_{34} + \beta\Omega\frac{\Omega + \gamma_r}{\Gamma_{12}} \tag{3}$$

with $\Gamma_{12} = \gamma_r + \gamma_{12} + \gamma'_{12}$. Here, it was assumed that excitation from MS$_2$ to MS$_3$ happens at a rate $\beta\Omega$. From fits of Eq. (3) on the extracted $\gamma'_{34}$ values [Fig. 4(b)], we obtain $\beta = 13.58 \pm 1.38\%$, which implies that excitation from MS$_2$ to MS$_3$ is approximately 7 times less efficient than from ground $^4A_2$ to first excited $^4A_2$ state. Furthermore, we find that $\gamma'^{(0)}_3 = 0.00 \pm 0.02$ MHz and $\gamma'^{(0)}_4 = 0.04 \pm 0.03$ MHz, which implies that almost no decay from MS$_2$ to the ground state happens intrinsically as a first-order transition. The observed decay is therefore mostly laser induced through excitation to MS$_3$, followed by subsequent decay to the first excited states and from there by radiative decay to the ground state spin manifolds. This is consistent with the scenario suggested by our ab initio level structure.

Strain causes similar intrinsic decay rates ($\gamma'^{(0)}_3 = 0.00 \pm 0.01$ MHz and $\gamma'^{(0)}_4 = 0.02 \pm 0.02$ MHz) and excitation happens slightly less efficient with $\beta = 0.11 \pm 0.02$. For this reason, we conclude that moderate strain does not affect the MS$_2$ dynamics.

**Fitting of experimental data with Lindblad Master Equation**

To obtain the radiative and non-radiative decay rates as depicted in Fig. 1(a) in combination with the power-dependent deshelving rates $\gamma'_{3,4}$, the time-evolution of the $V_{Si}$ was simulated through numerical solutions of the Lindblad master equation [10,12,29],

$$\begin{aligned}\dot{\rho} = &-i[H_{tot}, \rho] + \sum_{m_s} L^{rad.}\left[|e, m_s\rangle \rightarrow |g, m_s\rangle\right] + \sum_{m_s} L^{ISC}\left[|e, m_s\rangle \rightarrow |ms_{1,2}\rangle\right] \\ &+ \sum_{m_s} L^{ISC}\left[|ms_{1,2}\rangle \rightarrow |g, m_s\rangle\right] + \sum_{m_s} L^{730nm}\left[|g, m_s\rangle \leftrightarrow |e, m_s\rangle\right].\end{aligned} \tag{4}$$

Here the summation of the individual Lindblad terms is performed over all spin states, $m_s = \left\{+\frac{3}{2}, +\frac{1}{2}, -\frac{1}{2}, -\frac{3}{2}\right\}$. $H_{tot}$ is the total Hamiltonian of ground, excited and metastable states including the effect of resonant laser drives. Radiative decay from excited to ground states is incorporated via $L^{rad.}$ and $L^{ISC}$ describe the non-radiative metastable state transitions. Furthermore, we model the off-resonant laser at 730 nm as additional terms into the Lindblad master equation via $L^{730nm}$ [29]. More details about the full spin Hamiltonian in the rotating frame, as well as the fitting procedure can be found in SI 6, SI 7 and Tab. (SI) 5.

| Parameter | $\gamma_r$ | $\gamma_1$ | $\gamma'_1$ | $\gamma_2$ | $\gamma'_2$ | $\gamma_3$ | $\gamma_4$ |
|---|---|---|---|---|---|---|---|
| No strain (MHz) | $56.39^{+10.38}_{-8.57}$ | $83.11^{+18.26}_{-14.70}$ | $26.89^{+5.62}_{-4.88}$ | $6.70^{+1.50}_{-1.18}$ | $27.33^{+6.08}_{-4.78}$ | $3.81^{+0.83}_{-0.68}$ | $0.24^{+0.05}_{-0.04}$ |
| Strain (MHz) | $56.36^{+12.40}_{-10.03}$ | $84.15^{+18.06}_{-15.42}$ | $27.51^{+6.16}_{-4.98}$ | $6.73^{+1.47}_{-1.23}$ | $27.63^{+5.99}_{-5.01}$ | $1.11^{+0.24}_{-0.20}$ | $0.09^{+0.02}_{-0.02}$ |

*Table 2: Metastable parameters with and without strain.*

By fitting Eq. (2) to the pulse sequences from Fig. 2(c), Fig. 3 and Fig. 4 we obtain the metastable state rates as displayed in Tab. 2. The rates into the metastable states, i.e. $\gamma_r$, $\gamma_{1,2}$ and $\gamma'_{1,2}$ only exhibit minor strain dependence, which is a direct consequence of the similar excited state lifetimes. This is in contrast to $\gamma_3$ and $\gamma_4$, which imply a strain-induced increase of the lifetime of $MS_1$ from 247 ns to 450 ns.

**Discussion**

In order to gain insight into the strain-dependence of the direct intersystem crossing rates (Tab. (SI) 3) between the first excited quartet $^4A'_2$ and the upper group of intermediate metastable states $^2A_1, {}^2E''$ as well as the lower group of intermediate states $^2A_2, {}^2E$ as shown in Fig. 1(a), we have analysed the strain-dependence of the contributions to the corresponding ISC rate for uniaxial strain along c-, a- and m-axis, namely the excitation energy, spin-orbit coupling, and the vibrational overlap (lineshape) - see section SI2 in the supporting information. As demonstrated by Figs. SI3-SI5, neither of these parameters show relevant strain-dependence for the strain range compatible with the experimentally observed 1.35 µeV strain-induced shift of the $V_{Si}$ center PL-line. This translates into only slightly modified direct rate constants, unless the Jahn-Teller dynamics in the corresponding doublet states plays a significant role. The small strain-independence of $\gamma_{1,2}$ and $\gamma'_{1,2}$, which are based on the ISC transitions involving $^2A_1, {}^2E''$ is compatible with this finding and in agreement with our calculated rates as shown in Tab. (SI) 2.

The rates $\gamma_{3,4}$, on the other hand, show a strain-induced reduction by a factor ~2.7-3.4. In order to understand this the following points are important: (i) the calculated rates for the transition from the vibrational ground state of $^2E$ to the quartet groundstate is much slower than the rates $\gamma_{3,4}$ even without including motional quenching, indicating that the transition must take place via a vibrationally excited $^2E$ state, (ii) the strain Hamiltonian indicates that symmetry lowering strain along x- and y-axis is present in the strained defect, which modifies the Jahn-Teller dynamics by distorting and tilting the potential energy surface along the strain field.

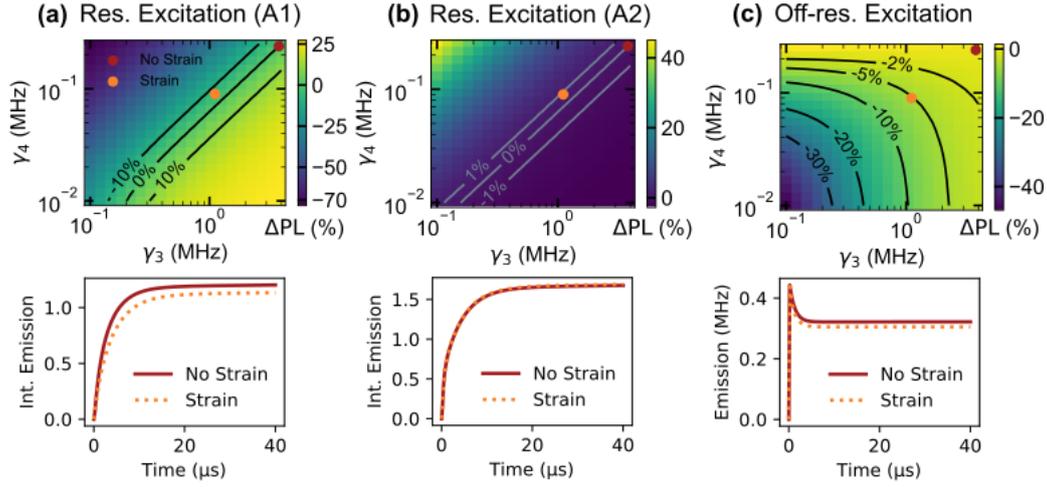

**Figure 5:** Simulated $V_{Si}$ emission properties in dependence of $\gamma_3$ and $\gamma_4$. (a) Relative change of integrated photoluminescence emission upon $40\ \mu s$ long resonant excitation along $A_1$, with respect to the photoluminescence as obtained for the unstrained color center. Lower plot: Simulated integrated emission of both color centers for durations from $0-40\ \mu s$. (b) Same simulation as in (a), except that resonant excitation is performed along $A_2$. 916 nm Rabi frequencies in (a) and (b) are 4.33MHz, corresponding to 20 nW of excitation power. Lower panels of (a) and (b) are simulations of the integrated emissions with the $\gamma_{3,4}$ as derived for, respectively, strain and no-strain. (c) Relative change of photoluminescence emission upon $40\mu s$ long off-resonant excitation with $\Omega_{730nm}=1\ MHz$, revealing a 5 % reduced count rate due to slower decay from $MS_1$ to the ground states.

We would like to conclude our study by investigating altered $V_{Si}$ emission properties, being subject to strain-induced changes of $\gamma_3$ and $\gamma_4$. Based on the present study we are not able to predict the exact changes in $\gamma_{3,4}$ as a function of the strain exerted on the 4H-SiC lattice. Nevertheless, estimates can be made by simulating the emission properties, depending on $\gamma_3$ and $\gamma_4$. As we find a predominant reduction of $\gamma_3$ [see Tab. 2] which describes non-radiative decay from $MS_1$ to $|g,\pm 1/2\rangle$, especially resonant excitation along $A_1$ is susceptible to changes of the decay rates [Fig. 5(a)]. As slower transition rates out of $MS_1$ implicate higher total cycling durations, the total emission is reduced and for $A_1$ excitation we expect a reduction of $\approx -5.7\%$ with respect to the unstrained case. This reduction is less pronounced if the $V_{Si}$ is excited with a 916nm laser resonant with the $A_2$ transition and we hereby expect changes $< 1\%$ [Fig. 5(b)]. Moreover, a reduced photoluminescence will be present upon off-resonant excitation [Fig. 5(c)] where relatively strong reductions are possible with reduced transition rates. In agreement with the findings for resonant excitation along $A_1$ and $A_2$ we expect a reduced emission signal of -5.1% for the strain configuration present within this study. As discussed in SI10, this is in accordance with cw measurements of the emitted photoluminescence under off-resonant 730 nm illumination.

**Conclusion & Outlook**

In conclusion, we have presented a comprehensive description of the strain-induced spin dynamics of $V_{Si}$ centers in 4H-SiC. Using a qualitative strain Hamiltonian, we are able to identify both axial and transverse strain contributions, revealing that transitions from $MS_1$ to the ground-state quartets are notably suppressed. As our methodology relies exclusively on all-optical pulse sequences, this approach enables strain characterization in regimes where microwave driving is not feasible—for instance, at millikelvin temperatures, where conventional microwave structures lead to excessive heating, or in nanophotonic devices that are not located near microwave antennas.

**Author Contributions**

M. H. performed all experiments. M. H., F. D. H. and R. N. designed the experiments. M. H., K. U. and S. K. P. prepared the experimental apparatus. D. S. and W. K. prepared the sample. M. H., R. N. and D. B. R. D. analyzed the data. M. S., M. N., and M. B. developed the ab initio methodology and performed the first-principles calculations. R. N. supervised the project. All authors contributed to the manuscript and provided essential input.

**Data Availability**

The data presented in this manuscript is available from the corresponding author on reasonable request.

**Acknowledgement**

We thank Florian Kaiser (LIST Luxembourg) for fruitful discussions. This work was supported by the European Union under the Key Digital Technologies Joint Undertaking (KDT JU)/Chip Joint Undertaking (Chips JU) in the projects ARCHIMEDES (grant no. 16MEE0329, 101112295) and MOSAIC (grant no.16MEE0494, 101081238) and by the Bundesministerium für Forschung, Technologie und Raumfahrt (BMFTR) in the projects INNOBAT (grant no. 03XP0492D), QuaLiProM (grant no. 03XP0573C), and QMNDQNet (grant no. 13N16264). M.S., M. N., and M. B. acknowledge financial support from the Austrian Science Fund (FWF, grant I5195) and German Research Foundation (DFG, QuCoLiMa, SFB/TRR 306, Project No. 429529648). The project profited from very generous computer time provided by the Erlangen National High Performance Computing Center (NHR@FAU) of the Friedrich-Alexander-Universität Erlangen-Nürnberg (FAU) and the Austrian Scientific Computing (ASC). D. D. would like to acknowledge the support of BMFTR through project QECHQS (Grant No. 16KIS1590K).

# Supplementary Information

## SI1. Experimental details

Our measurements were performed in a home-built confocal microscope. For resonant excitation, a tunable diode laser at 916 nm was used (Toptica DL Pro) and off-resonant excitation was achieved with the Toptica iBeam smart at 730 nm. Pulsed operation of the DL Pro was achieved by coupling into an acousto-optical modulator (AOM) from Gooch & Housego. For pulse sequences where interleaved pumping along $A_1$ and $A_2$ was performed, the center frequency of the 916 nm laser was detuned from both transitions and sidebands at the desired transitions were created with an electro-optical modulator (EOM, Exail MPX-950-LN-10). The drive frequencies of the EOM were created with both a Windfreak and an AWG from Zurich Instruments (HDAWG). To filter the laser photons from the $V_{Si}$ phonon-sideband emission, a 950 nm longpass filter from Thorlabs (FELH0950) was used. The sideband emission was coupled into optical fibers and fed to a superconducting nanowire single photon detector (SNSPD) from Single Quantum. The sample was placed inside a Qinu cryostat, operated at 4K. Excitation of the sample and collection of the emitted fluorescence was performed with a high-NA objective from Attocube (LT-APO).

The detected photons were time-tagged with a Swabian Instruments Time Tagger Ultra. Both the execution of the respective pulses as well as the post-processing of the time-tagged photons were orchestrated with LabOneQ and the Zurich Instruments HDAWG.

## SI2. Ab initio theory

### SI2.1. Ab initio calculations of energy levels of the $k$-$V_{Si}$ in 4H SiC

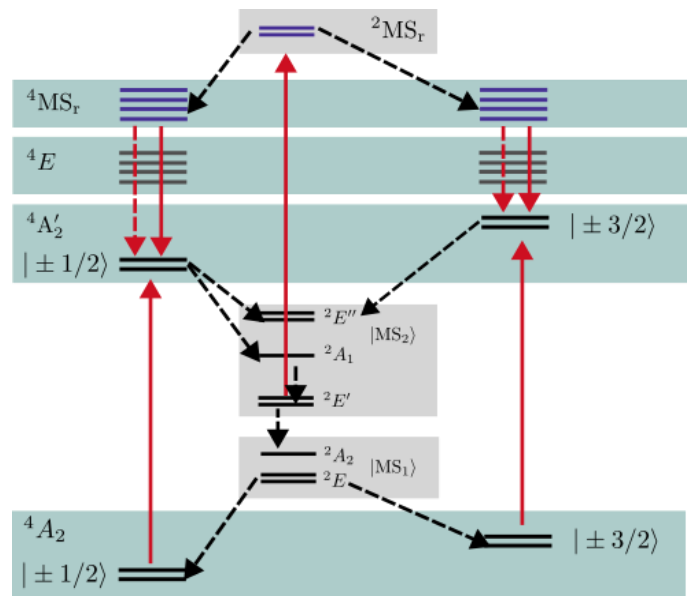

***Figure (SI) 1***: *Full k-$V_{Si}$ electronic structure CI-cRPA method. Non-radiative first order transitions are indicated by black dashed arrows. Optical/Vibrational transitions are indicated as solid/dashed red arrows. Consistently with previous studies [21], our rate equation model [Fig. 1(a), main text] does include transitions from $MS_2$ to $MS_1$ via $^2E'$, but instead fits the second-order transitions from $MS_2$ to the ground state quartets [see Fig. 1(b), main text]. Similarly, transitions from the excited states to $MS_1$ via $MS_2$ and $^2E'$ are taken into account through the second-order $\gamma_{1,2}$ rates.*

We calculate the full electronic multiplet structure of $k$-$V_{Si}$ using a combined framework of density functional theory and the many-body embedding CI-cRPA method [30]. Equilibrium structures for the ground- and first excited electronic state are obtained using spin-polarised density functional theory alongside the HSE06 hybrid functional [31], as implemented in the VASP software suite [32]. Large 576-atom supercells are used in order to reduce finite size effects, allowing us to sample the reciprocal space exclusively at the $\Gamma$-point. The relaxed geometries of the ground and excited state quartets $^4A_2$ and $^4A'_2$ were obtained via relaxation within DFT and CDFT schemes, respectively. As outlined in Ref. [30] we then construct an embedded many-body Hamiltonian

$$H_{CI-cRPA} = \sum_{ij}(\epsilon_i \delta_{ij} - h_{ij}^{DC}) a_i^\dagger a_j + \sum_{ijmn} \langle ij|V_{eff}|nm\rangle a^\dagger_i a^\dagger_j a_m a_n \quad (SI\ 1)$$

using a set of spin-restricted Kohn-Sham states $|i\rangle$ and -energies $\epsilon_i$ comprised of the localized defect states, resonances, and valence band states in the energy range extending down to 1.2eV below the valence band edge, and an effective screened electron-electron interaction $V_{eff} = \varepsilon^{-1}_{cRPA} V_{Coulomb}$ calculated using the constrained Random Phase Approximation (cRPA). Additionally, we correct for double counting by adding $h_{ij}^{DC}$ to the one-particle term as detailed in [30]. In order to obtain the excited state energy at a geometry $Q$ along the linear path between the $^4A_2$ and $^4A_2'$ geometries, the CI-cRPA vertical multiplet excitation energies at $Q$ are added to the energy difference $\Delta = E^{(0)}(Q) - E^0(0)$, where $E^{(0)}(Q)$ is the DFT ground-state energy at $Q$. Our multiplet energies in Tab. (SI) 1 agree with earlier reported CI-cRPA values [22,23] using slightly different lattice constants.

| State | Geometry | | |
|---|---|---|---|
| | Ground State | 1st Exc. State | Interp. Min. |
| $^4A_2$ | 0 | 0.13 | 0 |
| $^2E$ | 0.25 | 0.31 | 0.25 |
| $^2A_2$ | 0.28 | 0.44 | 0.28 |
| $^2E'$ | 0.87 | 0.88 | 0.84 |
| $^2A_1$ | 1.14 | 1.28 | 1.14 |
| $^2E''$ | 1.14 | 1.28 | 1.14 |
| $^4A'_2$ | 1.38 | 1.19 | 1.19 |
| $^4E$ | 1.48 | 1.53 | 1.46 |
| $^2MS_r$ | 2.19-2.26 | 2.01-2.06 | - |
| $^4MS_r$ | 2.10-2.24 | 2.01-2.21 | - |

*Table (SI) 1: Multiplet Energies in eV obtained using the ground $^4A_2$ and first excited $^4A'_2$ state geometries, as well as interpolated minima of the intermediate states along the linear path between the $^4A_2$ and $^4A'_2$ geometries.*

Besides the intermediate multiplet states, we identify a range of excitonic doublet $^2MS_r$ and quartet $^4MS_r$ states. Notably, $^2MS_r$ exhibits remarkably strong coupling (see Table (SI 2)) to the $^2E'$ state in the same energetic window as the $^4A_2 \rightarrow {}^4A'_2$ vertical excitation (red arrows in Fig. SI1). Analogously the transition dipole moment between $^4MS_r$ the excited quartets states is similarly strong. Given their energetic proximity, non-radiative crossing between these excitonic states therefore could open alternative coupling paths restoring $^4A'_2$ after initial ISC into the doublet manifold (see Fig. (SI)1)

| | $^4A_2 \to {}^4A'_2$ | $^2E' \to {}^2MS_r$ | $^4MS_r \to {}^4E$ | $^4MS_r \to {}^4A'_2$ |
|---|---|---|---|---|
| $|\vec{d}_T| / |\vec{d}_{A_1}|$ | 1 | 1-8 | 1-22 | 14-32 |

**Table (SI) 2**: Transition Dipole magnitudes $|\vec{d}_T|$ of $^4MS_r$ and $^2MS_r$ transitions in terms of those of A1/A2 $|\vec{d}_{A_1}|$.

### SI 2.2. Spin-Orbit Mediated Intersystem Crossing

Intersystem crossing (dashed lines in Fig. (SI)1) between quartets and doublets requires transitions $\Delta S > 0$ disallowed by ordinary optical selection rules. Instead, it is mediated by Spin-Orbit coupling $\lambda_{SOC} = |\langle f|H_{SOC}|i\rangle|$ between the initial (*i*) and final (*f*) state and phonons following in accordance with Fermi's Golden Rule [33]

$$\Gamma^{ISC} = \frac{2\pi}{\hbar}\lambda_{SOC}^2 A(\Delta_{if}) \qquad (SI\ 2)$$

where $A(\Delta_{if})$ is the vibrational overlap (see SI 2.2) and $\Delta_{if}$ is the energy difference between the respective minima of $|i\rangle$ and $|f\rangle$. We estimate the first order ISC rates by calculating $\lambda^{SOC}$ for all relevant multiplet states under the approximation that SOC does not depend on $Q$ (Condon approximation). However, for the evaluation of $A(\Delta_{if})$ we still require the respective energetic minima and associated geometries. In case of the quartet states, they can be obtained by a CDFT relaxation scheme outlined in Section SI2.1. This does not apply for the highly correlated intermediate states of the silicon vacancy, for which relaxation within an CI-cRPA-based scheme is required. As such a scheme presently is unavailable, we bridge this gap by first interpolating the multiplet energies along the linear path between the ground and excited $^4A'_2$ state, and then inferring the minima from the resulting Born-Oppenheimer surfaces (see Fig. (SI) 2(b)). Spin-orbit matrix elements are calculated within the CI-cRPA scheme in a perturbative approach as outlined in [34] using the same basis of Kohn-Sham states as for the electronic structure calculations.

The resulting upper-branch rates given in Tab. (SI) 2 are approximately one order of magnitude faster compared to respective combined fitted rates $(\gamma_1 + \gamma'_1, \gamma_2 + \gamma'_2)$ depopulating $^4A'_2$. The source of this discrepancy likely lies in our neglect of orbital quenching of spin-orbit coupling by the pseudo- and dynamic Jahn-Teller coupling between $^2A_1$ and $^2E''$ as well as the sublevels of $^2E''$. In comparison to these transitions, the ISC rates $^4A'_2 \to {}^2E'$ are negligibly small due to the large energy separation of 350 meV between these states. We conclude that populating $^2E'$ is dominated by potentially relatively fast non-radiative coupling from ($^2E'', {}^2A_1$) or a higher order ISC process involving ($^2E'', {}^2A_1$). The magnitudes and spin-selectivity of our calculated doublet → quartet transitions deviate from experimental rates by multiple orders of magnitude. These transitions likely involve multi-phonon processes as well as quenching of SOC due to the dynamic Jahn-Teller processes which we do not include in our picture.

| ISC Process | Spin Flipping Rate (MHz) | Spin conserving Rate (MHz) |
|---|---|---|
| $^4A'_2 \to {}^2E''$ | 108 | 36 |
| $^4A'_2 \to {}^2A_1$ | ≈0 | 456 |
| $^4A'_2 \to {}^2E'$ | 0.3 | 1 |
| $^2E' \rightarrow {}^4A_2$ | 0.018 | 0.006 |
| $^2E \rightarrow {}^4A_2$ | 0.251 | 0.083 |

Table (SI) 3: First order Intersystem Crossing Rates obtained by ab-initio theory.

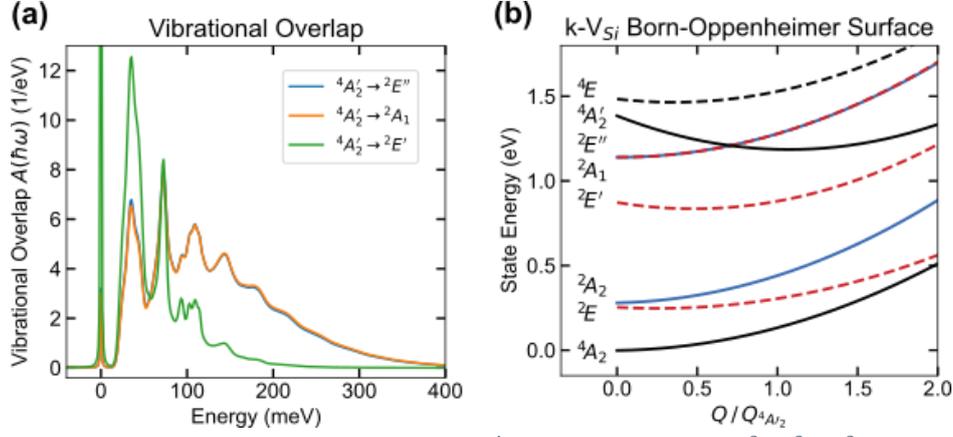

Figure (SI) 2: (a) Vibrational overlap functions for coupling the $^4A'_2$ excited state with $^2E''$, $^2A_1$, $^2E'$ doublets. The values for $A(\Delta)$ entering Eq. (SI 2) are calculated using the energy differences between the respective minima. (b) Born-Oppenheimer surfaces along the $^4A'_2$ effective displacement direction $Q_{^4A'_2}$.

### SI 2.3. Lineshape Function and Vibrational Overlap

Using the generating function approach [35], the normalized vibrational overlap function in Eq. (SI 2) can be written as

$$A(\hbar\omega) = \frac{1}{2\pi} \int dt \; exp\left(S(t) - S(0)\right) e^{-i\omega t - \eta|t|} \tag{SI 3}$$

where S(t) is the spectral density of electron-phonon coupling in the time domain and $\eta$ is a small factor chosen to reproduce the broadening of the transition. We obtain S(t) by Fourier transforming its frequency-dependent counterpart

$$S(\hbar\omega) = \sum_k S_k \, \delta(\hbar\omega - \hbar\omega_k) \approx \sum_k \frac{S_k}{\sqrt{2\pi}\sigma_k} e^{(\hbar\omega - \hbar\omega_k)^2/2\sigma_k^2} \tag{SI 4}$$

where the sum runs over all phonon modes, with $\omega_k$ being their respective frequencies, and $\sigma_k$ is a parameter to model the broadening of the individual modes. The partial Huang-Rhys factors $S_k$ are defined by

$$S_k = \frac{q^2_k \, \omega_k}{2\,\hbar} \tag{SI 5}$$

with $q_k$ being mass-weighted displacements

$$q_k = \sum_a \sqrt{m_a} \; \vec{R}_{if,a} \, \vec{e}_{k,a} \tag{SI 6}$$

where $\vec{R}_{if,a} = \vec{R}_{i,a} - \vec{R}_{f,a}$ is the displacement an atom ($a$) experiences between the initial and final defect geometries (e.g. ground and excited state equilibrium structures), $m_a$ is its mass, and $e_{k,a}$ represents its displacement in the k$^{th}$ phonon mode.

We calculate the phonon spectrum by employing the PHONOPY package [36]. Similar to our electronic calculations, we use 576-atom supercells sampled only at the $\Gamma$-point. Due to the large computational effort required by the reference calculations for the finite displacement method, we use the less expensive PBE functional. To calculate the lineshapes for quartet-doublet transitions shown in Fig. (SI) 2, we project the resulting PBE-modes onto the (interpolated) displacements obtained from our HSE calculations (see SI2) as outlined in Eq. (SI 6).

**SI 2.4. Strain-dependence of quartet and doublet levels and spin-orbit coupling**

The strain-dependence of the intersystem crossing rate $\Gamma^{ISC}$

$$\Gamma^{ISC} = \frac{2\pi}{\hbar}\lambda^2_{SOC}\ A(\Delta_{if}) \qquad (SI\ 7)$$

can be traced back to the strain dependence of the spin-orbit coupling matrix element, the transition energy $\Delta_{if}$, and of the vibrational overlap $A$.

The dependence of the quartet and doublet levels as well as the spin-orbit coupling matrix elements is calculated for uniaxial strain along the c-, a- and m- crystal directions (cf. Fig (SI) 3(a)) using CI-cRPA in combination with DFT. A strain range of [-0.15%, 0.15%] relevant for strained crystals is covered. For each strain value the defect geometry is fully relaxed. Fig. (SI) 3(b) show the excited quartet and intermediate doublet levels in the strained and relaxed ground state geometry vs. the applied strain. The calculated vertical excitation between the quartets is also shown. Within the range of strain values between -0.5% and +0.5% the calculated vertical excitation as well as the ZPL (not shown here) changes only by 8 meV (c-axis), 3 meV (m-axis), and 2 meV (a-axis). The experimentally observed ZPL-shift of 1.35 µeV indicates that much lower strain fields are present in the experiments. In case of strain along a- and m-directions the symmetry of the defect is lowered from $C_{3v}$ to $C_{1h}$, which leads to a splitting of the $^2E$ levels and a coupling to the corresponding $^2A_1$ or $^2A_2$ level in the upper and lower group of intermediate states. The strain-induced changes in the doublet levels typically vary by about 10meV across the range of strain values.

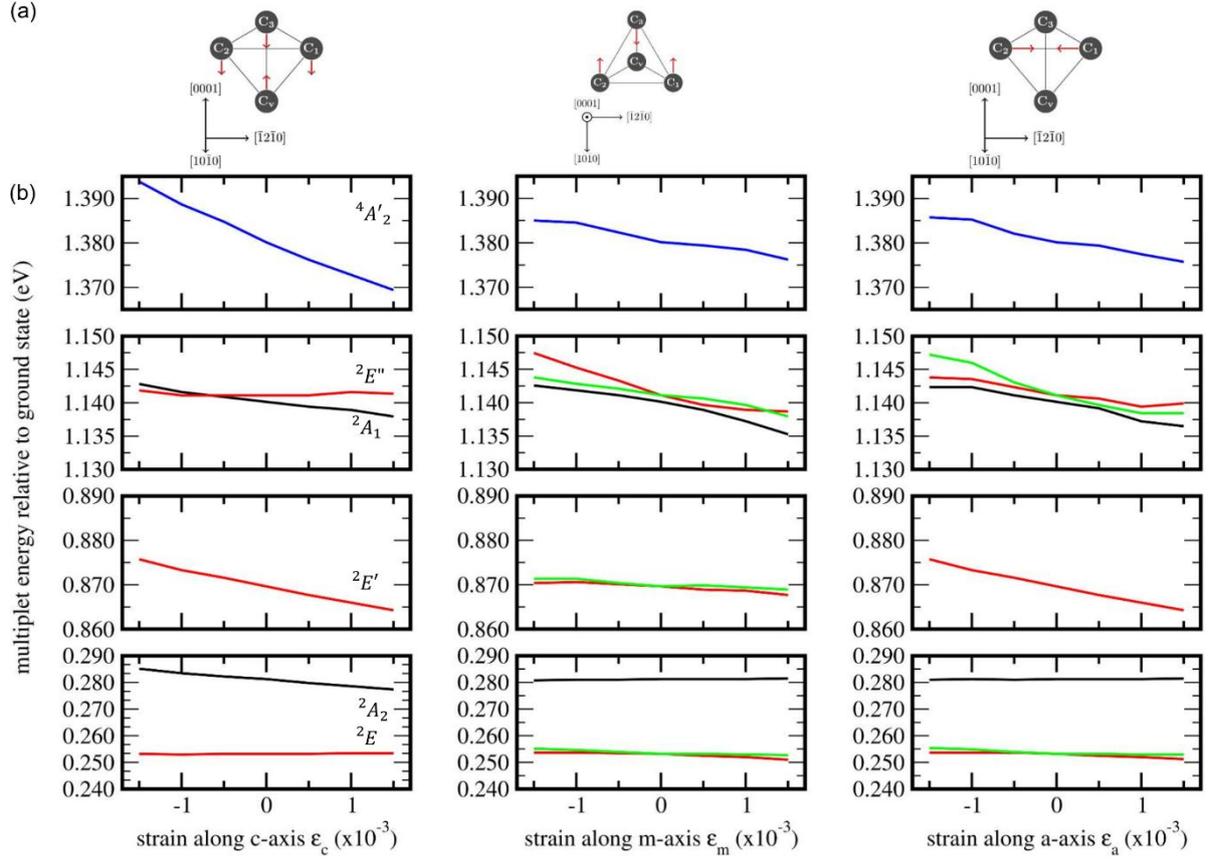

*Figure (SI) 3*: (a) Displacement of vacancy neighbors for uniaxial strain along the c-, m- and a-axis, respectively. Note the symmetry-lowering distortion in the latter two cases. (b) Energy of the excited quartet and groups of intermediate doublet states for the ground state geometry of the vacancy relative to the ground state energy vs. the applied uniaxial strain. Under the symmetry-lowering strain the degenerate levels $^2E$, $^2E'$, and $^2E''$ split into two non-degenerate levels of $^2A$ and $^2A'$ representations that are indicated by red and green lines in the corresponding panels.

The strain dependence of the spin-orbit coupling matrix elements is shown Fig. (SI) 3 both for the coupling of the intermediate doublet states to the relevant quartet states. We observe only a small variation of the values versus strain on the order of 5% in the range of strain values between -0.5% and 0.5%. The only exception occurs in the upper group of intermediate states $^2A_1$ and $^2E''$ for symmetry lowering tensile strain along the a- and m- direction. Here by strain-induced coupling between the $^2A_1$ and one partner of $^2E''$ in an avoided crossing of the spin-orbit coupling matrix elements between the states are exchanged. Given the close spacing of the levels, this however, should have little effect on the effective ISC rate involving this group of states and the energetic separations are not relevant.

Also for the $^2A_2$, we observe an increase of the spin-orbit coupling to the ground state for both symmetry lowering tensile and compressive strain to still negligible values compared to $^2E$.

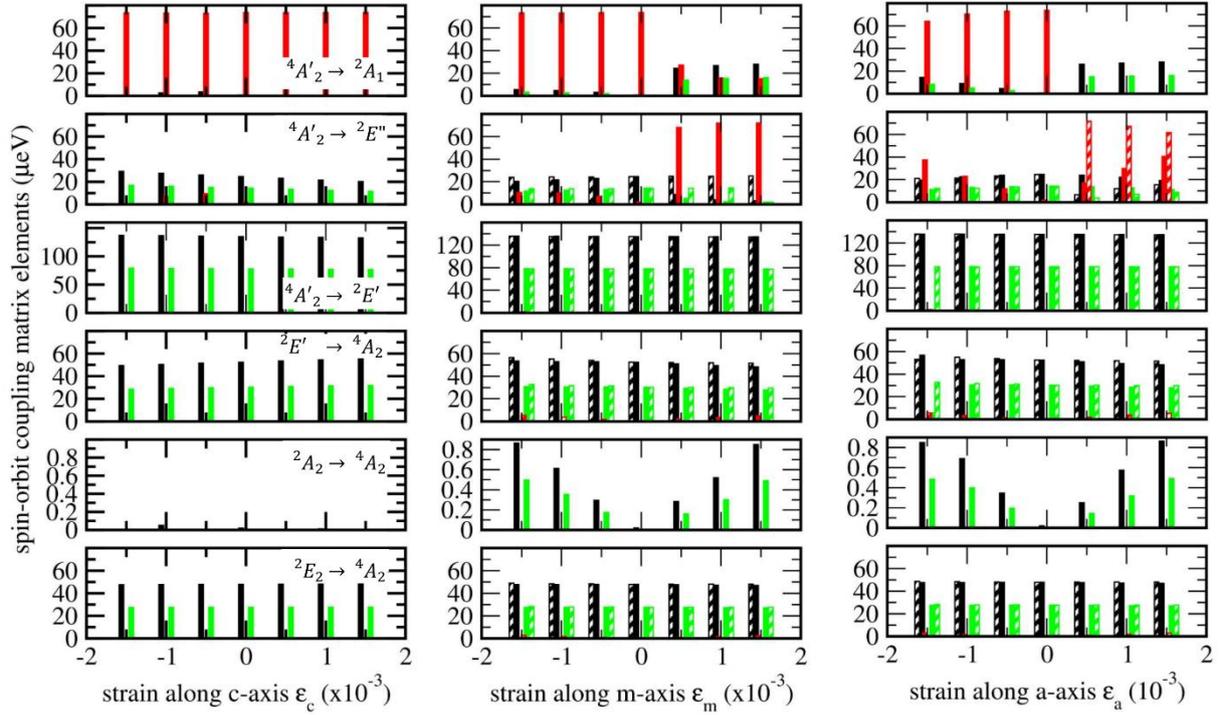

*Figure (SI) 4:* Spin-orbit coupling matrix elements vs. uniaxial strain along the c-, m, and a-axis for the ISC between the excited (ground) state quartet and the intermediate states indicated in the left panels. Black bars represent coupling between ±3/2 and ±1/2 spin states, red bars between +1/2 and +1/2 and green bars between +1/2 and -1/2. In case of the E-doublets with symmetry-lowering strain, the solid bars refer to the sublevel $E_x$ and the hatched bars to $E_y$. Note that the larger changes for symmetry lowering strain result from an avoided level crossing between the $^2A_1$ and one $^2E''$ state.

The strain dependence of the vibrational overlap in the range of strain relevant for our device also turns out to be small as shown exemplarily in Fig (SI) 5. Only for strain ~1% or larger significant changes are expected.

In summary, strain-induced changes in the energy levels, spin-orbit coupling strength, and the vibrational overlap should not lead to significant changes in the rates.

The strain-dependence of motional quenching in the $^2E$, $^2E'$ and $^2E''$ intermediate states, arising from a dynamic Jahn-Teller coupling, has not been considered here, as the detailed potential energy surfaces are not fully explored. The low strain dependence of the experimentally determined rates $\gamma_{1,2}$ and $\gamma'_{1,2}$ that we associate with the ISC to the group of states $^2A_1$, $^2E''$ indicates, however, that the strain-dependence of motional quenching is not relevant here.

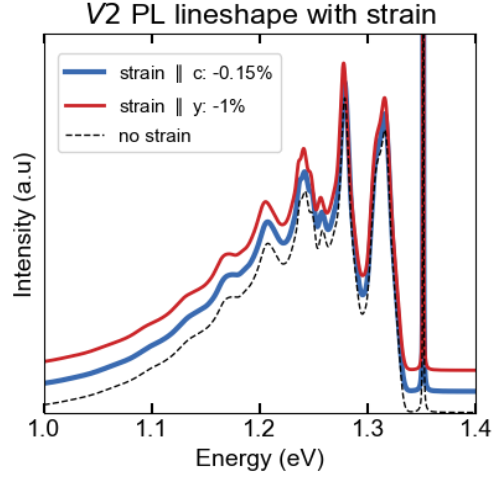

*Figure (SI) 5:* *Optical Lineshape (ZPL + PSB) for the V2 transition under different strain regimes. Lines corresponding to different strains are shifted against each other for visual clarity.*

### SI3. Spin-Strain Hamiltonian and C$_{3v}$ invariance

The most general form of a spin Hamiltonian, describing the influence of a strain tensor $\epsilon_{ij} = \frac{1}{2}\left(\frac{\partial u_i}{\partial x_j} + \frac{\partial u_j}{\partial x_i}\right)$ with $u(r)$ the displacement field of the crystal lattice was derived by [25] and reads

$$H_\epsilon = H_{\epsilon 0} + H_{\epsilon 1} + H_{\epsilon 2} \quad (SI\ 8a)$$

$$H_{\epsilon 0} = \left[h_{41}(\epsilon_{xx} + \epsilon_{yy}) + h_{43}\epsilon_{zz}\right]S_z^2 \quad (SI\ 8b)$$

$$H_{\epsilon 1} = \frac{1}{2}\left[h_{26}(\epsilon_{xx} + \epsilon_{yy}) + h_{43}\epsilon_{zz}\right]\{S_x, S_z\} + \frac{1}{2}(h_{26}\epsilon_{yz} + h_{25}\epsilon_{xy})\{S_y, S_z\} \quad (SI\ 8c)$$

$$H_{\epsilon 2} = \frac{1}{2}\left[h_{16}\epsilon_{zx} - \frac{1}{2}h_{15}(\epsilon_{xx} - \epsilon_{yy})\right](S_y^2 - S_x^2) + \frac{1}{2}(h_{16}\epsilon_{yz} + h_{15}\epsilon_{xy})\{S_x, S_y\} \quad (SI\ 8d)$$

Here $\{\cdot,\cdot\}$ denotes the anti-commutator. The coupling-strength parameters $h_{41}$, $h_{43}$, $h_{25}$, $h_{15}$ and $h_{16}$ are material specific and were derived for the NV center in diamond [25] and divacancy centers in 4H SiC [26]. We are not aware of any reported spin-strain coupling coefficients for cubic or hexagonal V$_{Si}$ centers in 4H SiC. Nevertheless, Eq. (SI8) can be re-expressed in terms of general strain components $\Pi$ as

$$H_{\epsilon 0} = \Pi_z S_z^2 \quad (SI\ 9a)$$

$$H_{\epsilon 1} = \Pi_1^{(1)}\{S_x, S_z\} + \Pi_2^{(1)}\{S_y, S_z\} \quad (SI\ 9b)$$

$$H_{\epsilon 2} = \Pi_1^{(2)}(S_y^2 - S_x^2) + \Pi_2^{(2)}\{S_x, S_y\}. \quad (SI\ 9c)$$

By introducing ladder operators $S_\pm = S_x \pm iS_y$ and writing $\Pi_{1,2}^{(1,2)}$ in polar coordinates,

$$\Pi_1^{(1,2)} = \Pi^{(1,2)} \cos\theta^{(1,2)}$$

$$\Pi_2^{(1,2)} = \Pi^{(1,2)} \sin\theta^{(1,2)} \quad (SI10)$$

Eq. (SI 9) can be compactly derived as

$$H_{\epsilon 0} = \Pi_z S_z^2 \quad (SI\ 11a)$$

$$H_{\epsilon 1} = \Pi^{(1)} \frac{e^{-i\theta^{(1)}}}{2}(S_+ S_z + S_z S_+) + \Pi^{(1)} \frac{e^{+i\theta^{(1)}}}{2}(S_- S_z + S_z S_-) \qquad (SI\ 11b)$$

$$H_{\epsilon 2} = \Pi^{(2)} \frac{e^{-i\theta^{(2)}}}{2} S_+^2 + \Pi^{(2)} \frac{e^{+i\theta^{(2)}}}{2} S_-^2 . \qquad (SI\ 11c)$$

Eq. (SI 11) represents the full spin-strain interaction under $C_{3v}$ symmetry. However, this expression can be further simplified by requiring symmetry invariance. In particular rotations along the z-axis, described via

$$U_\phi = e^{-i\phi S_z} \qquad (SI\ 12)$$

transform Eq. (SI 11) towards

$$U_\phi H_{\epsilon 0} U_\phi^\dagger = \Pi_z S_z^2 \qquad (SI\ 13a)$$

$$U_\phi H_{\epsilon 1} U_\phi^\dagger = \Pi^{(1)} \frac{e^{-i(\theta^{(1)}+\phi)}}{2}(S_+ S_z + S_z S_+) + \Pi^{(1)} \frac{e^{+i(\theta^{(1)}+\phi)}}{2}(S_- S_z + S_z S_-) \qquad (SI\ 13b)$$

$$U_\phi H_{\epsilon 2} U_\phi^\dagger = \Pi^{(2)} \frac{e^{-i(\theta^{(2)}+2\phi)}}{2} S_+^2 + \Pi^{(2)} \frac{e^{+i(\theta^{(2)}+2\phi)}}{2} S_-^2 \qquad (SI\ 13c)$$

This means that $U_\phi$ effectively transforms $\theta^{(1)} \to \theta^{(1)} + \phi$ and $\theta^{(2)} \to \theta^{(2)} + 2\phi$. Nevertheless, one can calculate that $\theta^{(2)} - 2\theta^{(1)} \to \theta^{(2)} - 2\theta^{(1)}$ which means that only the difference between both phases $\theta^{(1,2)}$ is invariant, not their precise value. Therefore, we set $\theta^{(1)} = 0$ and keep only $\theta^{(2)} \equiv \theta$ in the following analysis.

Rotations by 180°, parametrized by

$$U_x = e^{-i\pi S_x} \qquad (SI\ 14)$$

Lead to

$$U_x H_{\epsilon 0} U_x^\dagger = \Pi_z S_z^2 \qquad (SI\ 14a)$$

$$U_x H_{\epsilon 1} U_x^\dagger = \Pi^{(1)} \frac{e^{+i\theta^{(1)}}}{2}(S_+ S_z + S_z S_+) + \Pi^{(1)} \frac{e^{-i\theta^{(1)}}}{2}(S_- S_z + S_z S_-) \qquad (SI\ 14b)$$

$$H_{\epsilon 2} = \Pi^{(2)} \frac{e^{+i\theta^{(2)}}}{2} S_+^2 + \Pi^{(2)} \frac{e^{-i\theta^{(2)}}}{2} S_-^2 . \qquad (SI\ 14c)$$

$U_x$ therefore effectively transforms $\theta^{(1,2)} \to -\theta^{(1,2)}$, which is similar to $\theta^{(1,2)} \to \theta^{(1,2)} + \pi$ as the transformation exchanges complex conjugates. In other words, terms of the form of $e^{\pm i\theta}$ transform as $-e^{\pm i\theta}$, which is equivalent to $\Pi^{(1,2)} \to -\Pi^{(1,2)}$. For this reason, only the magnitude of $\Pi^{(1,2)}$ can be determined from Eq. (SI 11), not its particular sign. The parameter ranges which we allow for the fitting of the spin-strain parameters as described in the main text are therefore $\Pi_z \in \mathbb{R}$, $\Pi^{(1,2)} \geq 0$, $\theta \in [0, \pi)$.

### SI4. Optical measurements of spin initialization fidelity

The ground state spin contrast can be defined as

$$C = \frac{\left|p_{\pm\frac{3}{2}} - p_{\pm\frac{1}{2}}\right|}{p_{\pm\frac{3}{2}} + p_{\pm\frac{1}{2}}}. \qquad (SI\ 15)$$

Here $p_{\pm\frac{1}{2}}$ and $p_{\pm\frac{3}{2}}$ represent the probabilities of finding the system in $|g,\pm 1/2\rangle$ or $|g,\pm 3/2\rangle$, respectively. These probabilities can be probed by applying short (300 ns) laser pulses along A$_1$ or A$_2$, which allows to re-write the ground state spin contrast in terms of the integrated photoluminescence signals as

$$C = \frac{\left|PL_{A_2} - PL_{A_1}\right|}{PL_{A_2} + PL_{A_1}}. \qquad (SI\ 16)$$

The measured PL signals are the result of both the emitted V$_{Si}$ photons and the detector dark counts ($PL_{dark}$). For this reason, one has

$$C_{ideal} = \frac{\left|(PL_{A_2} - PL_{dark}) - (PL_{A_1} - PL_{dark})\right|}{(PL_{A_2} - PL_{dark}) + (PL_{A_1} - PL_{dark})}. \qquad (SI\ 17)$$

With this, the ground state spin contrast after 80 $\mu s$ illumination with the resonant laser at A$_1$ and 7 Hz dark count rate is $C_{ideal} = 98.04 \pm 2.53\ \%$ %, where the error has been estimated from the Poissonian distribution of the collected photons [Fig. (SI) 6].

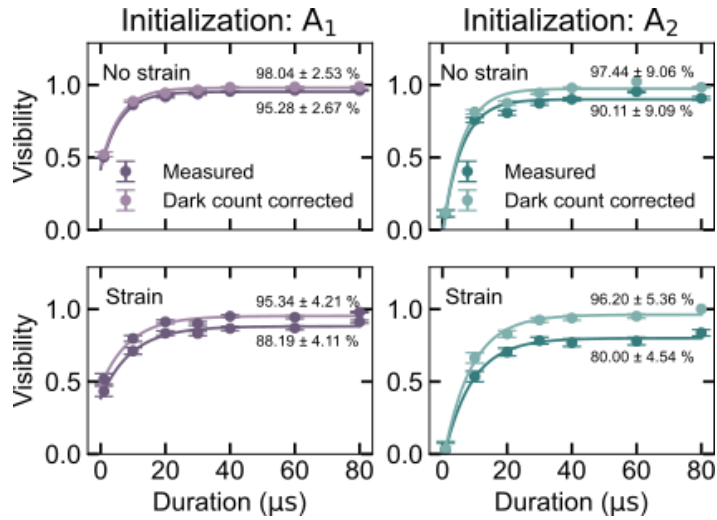

*Figure (SI) 6: Optically detected ground state visibility after polarizing the V$_{Si}$ with 20 nW laser power and the resonance tuned either at A$_1$ (left column) or A$_2$ (right column). Dark curves display the contrast as inferred from Eq. (SI 9) and lighter curves include background correction of 7 Hz detector dark counts [Eq. (SI) 17]. Upper row are visibility measurements in absence of strain and lower row are the results for the strained V$_{Si}$. Even if dark count correction is included, strain causes visibility reduction by $\approx 2-3\%$.*

As one can see from Fig. SI6, strain causes reduction of the dark count corrected ground state visibilities of $98.04 \pm 2.54\% \rightarrow 95.34 \pm 4.21\%$ for initialization with A$_1$ laser light and $97.44 \pm 9.06\% \rightarrow 96.20 \pm 5.36\%$ for A$_2$.

**SI5. Ground state spin polarization by 730nm laser**

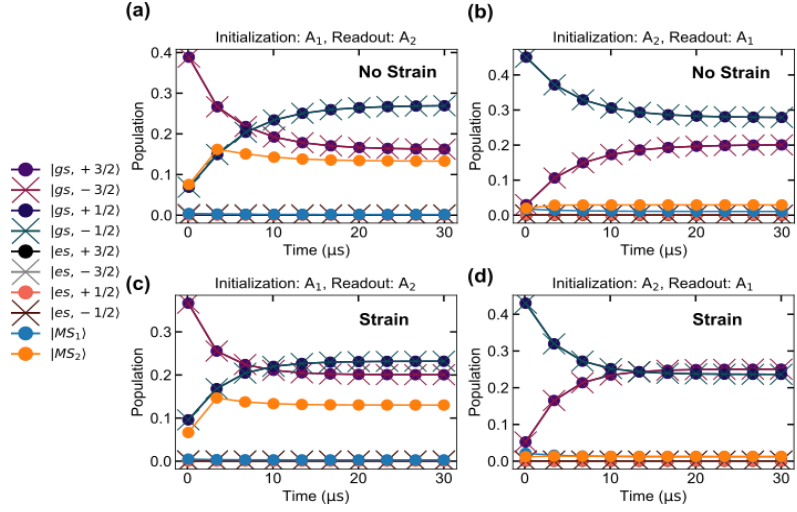

*Figure (SI) 7: Fitted time-evolution of the $V_{Si}$ spin states during the spin depolarization under 730 nm laser drive. The populations are inferred from the fit results of Fig. 3(b) (main text).*

To obtain the ground state spin polarization under continuous laser drive, we extract the $V_{Si}$ populations from the Lindblad fit as displayed in Fig. 3(b) of the main text [Fig. SI 7]. The ground state polarizations are then given as

$$P_{\pm\frac{1}{2}} = \frac{p^g_{+\frac{1}{2}} + p^g_{-\frac{1}{2}}}{p^g_{+\frac{3}{2}} + p^g_{-\frac{3}{2}} + p^g_{+\frac{1}{2}} + p^g_{-\frac{1}{2}}} \qquad (SI\ 18)$$

and

$$P_{\pm\frac{3}{2}} = \frac{p^g_{+\frac{3}{2}} + p^g_{-\frac{3}{2}}}{p^g_{+\frac{3}{2}} + p^g_{-\frac{3}{2}} + p^g_{+\frac{1}{2}} + p^g_{-\frac{1}{2}}}. \qquad (SI\ 19)$$

From the populations as shown in Fig. SI 7 the ground state spin polarizations after 30 μs of continuous 730 nm laser drive (power 50 μW) are shown in Tab. (SI) 4. In particular, we find that in absence of strain the ground state spin polarization under continuous 730 nm laser drive is $\frac{P_{\pm 1/2}}{P_{\pm 3/2}} = \frac{60.28\%}{37.72\%}$. The strain as present in this study reduces this polarization to $\frac{52.53\%}{47.47\%}$.

|  |  | Initialization: A1, Readout: A2 | Initialization: A2, Readout: A1 | Average |
|---|---|---|---|---|
| No Strain | $P_{\pm 1/2}$ | 62.39 % | 58.18 % | 60.28 % |
|  | $P_{\pm 3/2}$ | 37.61 % | 41.82 % | 39.72 % |
| Strain | $P_{\pm 1/2}$ | 53.62 % | 51.44 % | 52.53 % |
|  | $P_{\pm 3/2}$ | 46.38 % | 48.56% | 47.47% |

*Table (SI) 4: Extracted ground state spin polarizations as obtained from the Lindblad Fits of Fig. 3(b) (main text).*

### SI6. Full $V_{Si}$ Hamiltonian in rotating frame

In the following, the full Hamiltonian of the V$_{Si}$, consisting of ground- and excited states will be derived. If ground and excited states are described by spin Hamiltonians $H_g$ and $H_e$, respectively, then the full Hamiltonian in the lab frame is

$$H_0 = |g\rangle\langle g| \otimes H_g + |e\rangle\langle e| \otimes (H_e + 1_4 \cdot \omega_{ZPL})$$

$$H_L(t) = \frac{\Omega_L}{2}\left(e^{i\omega_L t}|g\rangle\langle e| + h.c.\right)$$

$$H_{tot} = H_0 + H_L(t). \tag{SI 20}$$

Here $|g\rangle\langle g|$ and $|e\rangle\langle e|$ denote ground and excited state subspaces, $\omega_{ZPL}$ is the zero-phonon-line frequency and laser drive with optical Rabi frequency $\Omega_L$ and frequency $\omega_L$ is incorporated via $H_L(t)$. Eq. (SI 20) can be transformed into a frame rotating with the laser frequency via the unitary transformation

$$U(t) = e^{i\omega_L t|e\rangle\langle e|\otimes 1_4}. \tag{SI 21}$$

The Hamiltonian in the rotating frame is then given as $H_{rot} = UHU^\dagger - iU\frac{d}{dt}U^\dagger$, which from direct calculation follows to be

$$H_{rot} = |e\rangle\langle e| \otimes (H_e + \delta_L 1_4) + |g\rangle\langle g| \otimes H_g + \frac{\Omega_L}{2}(|g\rangle\langle e| \otimes 1_4 + h.c.) \tag{SI 22}$$

Here $\delta_L = \omega_{ZPL} - \omega_L$ is the detuning between zero-photon line and laser frequency.

**SI7. Fits through Lindblad Master Equation**

To infer the rates in and out of the metastable states as well as the radiative decay rate, the time evolution of the V$_{Si}$ center was simulated by the Lindblad master equation,

$$\dot{\rho} = -i[H_{tot}, \rho] + \sum_{m_s=\{\pm\frac{1}{2},\pm\frac{3}{2}\}} L[|e, m_s\rangle \to |g, m_s\rangle] + \sum_{m_s=\{\pm\frac{1}{2},\pm\frac{3}{2}\}} L[\{|e, m_s\rangle\}_{m_s} \to \{|ms_2\rangle, |ms_1\rangle\}]$$

$$+ \sum_{m_s=\{\pm\frac{1}{2},\pm\frac{3}{2}\}} L[\{|ms_2\rangle, |ms_1\rangle\} \to \{|g, m_s\rangle\}_{m_s}] \tag{SI 23}$$

in units of $\hbar = 1$. Here $H_{tot}$ is the full V$_{Si}$ Hamiltonian, consisting of excited, ground and metastable states as well as laser drives. The Lindblad operators $L[|i\rangle \to |j\rangle]$ describe the transition from state $|i\rangle$ to $|j\rangle$ with an associated rate $\gamma$ as depicted in Fig. 1(a) (main text). For instance, one has

$$L\left[\left|e, m_s = \pm\frac{1}{2}\right\rangle \to |ms_2\rangle\right] = \gamma_1'\left[A\rho A^\dagger - \frac{1}{2}(\rho A^\dagger A + A^\dagger A\rho)\right] \tag{SI 24a}$$

$$A = |ms_2\rangle\left\langle e, m_s = \pm\frac{1}{2}\right| \tag{SI 24b}$$

All rates as depicted in Fig. 1(a) (main text) are added to the Lindblad master equation as terms similar to this expression. Besides this, we also model the off-resonant laser at 730nm as an incoherent process with Lindblad operators of the form of

$$\sqrt{\Omega_{730nm}}[|e, \pm 1/2\rangle\langle g, \pm 1/2| + |e, \pm 3/2\rangle\langle g, \pm 3/2| + h.c.]. \tag{SI 25}$$

The emitted (spontaneous) photoluminescence is then given in terms of the excited state matrix elements:

$$PL(t) = \eta \gamma_r \sum_{m_s=\{\pm\frac{1}{2},\pm\frac{3}{2}\}} \langle e, m_s | \rho(t) | e, m_s \rangle. \qquad (SI\ 26)$$

Here $\eta$ can be interpreted as the efficiency of the setup with its inverse being the detection efficiency of the emitted photons. To fit the rate equation model as depicted in Fig. 1(a) onto our experimentally measured data, the pulse sequences are fitted with help of Nelder-Mead optimization routines and the resulting parameters are listed in Tab. 2 (main text).

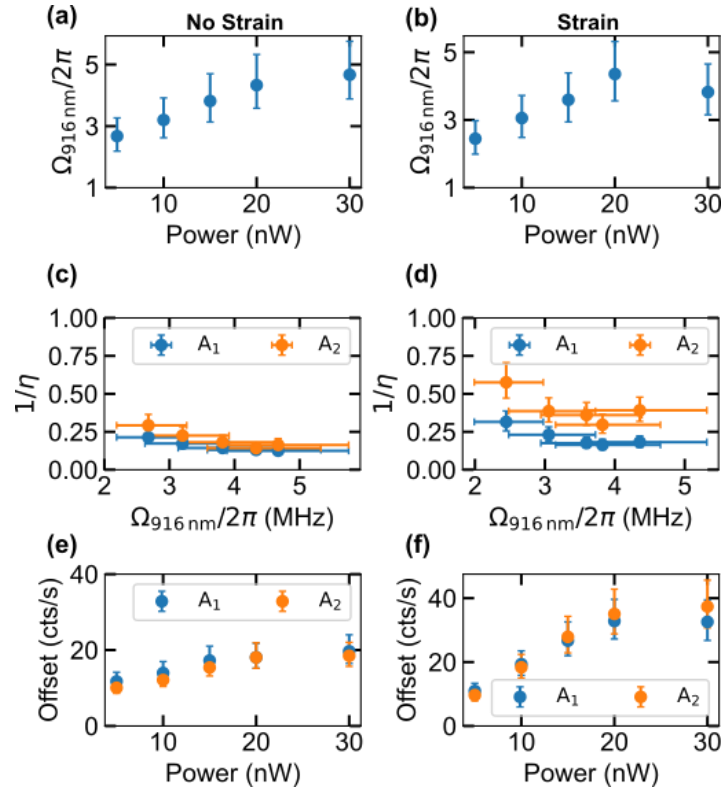

*Figure (SI) 8:* *Extracted 916 nm Rabi frequencies during the measurements as displayed in Fig. 4 of the main text of (a) no strain and (b) in presence of strain. The corresponding detection efficiencies of (c) no strain and (d) strain are lower than 1%. The offset during resonant spin depletion measurements of no strain and with strain are shown in, respectively, (e) and (f).*

From the fits as displayed in Fig. 3(a) of the main text, we also obtain the 916 nm Rabi frequencies which are displayed in Fig. (SI) 4(a, b). Furthermore, we obtain detection efficiencies which are below 1% [Fig. (SI) 4(c,d)]. We attribute these low values to finite coupling efficiencies into the SNSPD together with the absence of nanophotonic structures, leading to significant scattering within the SiC crystal. Furthermore, we allowed for the fitting of a small power dependent offset for the fits from Fig. 4(a) (main text) due to breakthrough through the used longpass filter and scattered laser light. It can be seen from Fig. (SI) 3(e,f), that the obtained values are in the order of $\sim 20$ cts/s.

During the fits, we also obtained values for the off-resonant laser drives, like the pumping strengths as well as the power-dependent deshelving rates of MS$_2$ during the off-resonant laser drives [Tab. (SI) 5].

| Power 730nm | $\Omega_{730nm}$ (MHz), No Strain | $\Omega_{730nm}$ (MHz), Strain | $\gamma'_3$ No Strain (MHz) | $\gamma'_4$ No Strain (MHz) | $\gamma'_3$ Strain (MHz) | $\gamma'_4$ Strain (MHz) |
|---|---|---|---|---|---|---|
| 50 µW | $0.60^{+0.13}_{-0.11}$ | $0.92^{+0.20}_{-0.17}$ | $0.18^{+0.04}_{-0.03}$ | $0.26^{+0.06}_{-0.05}$ | $0.17^{+0.04}_{-0.03}$ | $0.49^{+0.11}_{-0.09}$ |

| 815 µW | $13.95^{+3.01}_{-2.52}$ | $5.19^{+1.15}_{-0.19}$ | $0.25^{+0.06}_{-0.04}$ | $0.33^{+0.07}_{-0.06}$ | $0.05^{+0.01}_{-0.01}$ | $0.05^{+0.01}_{-0.01}$ |

*Table (SI) 5: Fit parameters for the off-resonant 730nm drive as obtained from the fits of the Lindblad master equation.*

### SI8. Approximate Bayesian Computation for Error Estimates

To obtain reliable error estimates, Approximate Bayesian Computation (ABC) is employed [27]. Hereby we start with the best-fit value as obtained from the Lindblad fits and allow variations of 20% from this value. For all fit parameters, i.e., all metastable-state rates, all Rabi frequencies, all efficiencies etc., a random set of values is chosen and the goodness of the fit is evaluated by calculation of $\chi_r^2$. If the value is within a confidence interval of 95%, then the parameters are saved, otherwise the randomly chosen set is discarded. This procedure is repeated 9000 times. From the resulting parameter distributions, the error of each fit parameter is obtained by numerically evaluating the 68% confidence interval.

### SI9. Power dependence of deshelving from MS$_2$

In the following, the power dependence of the rates from MS$_2$ to the ground state manifold will be derived based on a rate equation model of the V$_{Si}$ [Fig. 1(a), main text]. If the V$_{Si}$ is driven along A$_1$ ($\Omega_{A_1}$) or A$_2$ ($\Omega_{A_2}$), then the rate equations are

$$\dot{p}_{ms_3} = R p_{ms_2} - 2\Gamma p_{ms_3}$$

$$\dot{p}^e_{\pm 3/2} = \Gamma p_{ms_3} + \Omega_{A_2} p^g_{\pm 3/2} - (\Gamma_2 + \Omega_{A_2}) p^e_{\pm 3/2}$$

$$\dot{p}^e_{\pm 1/2} = \Gamma p_{ms_3} + \Omega_{A_1} p^g_{\pm 1/2} - (\Gamma_1 + \Omega_{A_1}) p^e_{\pm 3/2}$$

$$\dot{p}_{ms_2} = \gamma'_1 p^e_{\pm 1/2} + \gamma'_2 p^e_{\pm 3/2} - (R + \gamma'_{3,0} + \gamma'_{4,0}) p_{ms_2}$$

$$\dot{p}_{ms_1} = \gamma_1 p^e_{\pm 1/2} + \gamma_2 p^e_{\pm 3/2} - (\gamma_3 + \gamma_4) p_{ms_1}$$

$$\dot{p}^g_{\pm 3/2} = (\Omega_{A_2} + \gamma_r) p^e_{\pm 3/2} + \gamma_4 p_{ms_1} + \gamma'_{4,0} p_{ms_2} - \Omega_{A_2} p^g_{\pm 3/2}$$

$$\dot{p}^g_{\pm \frac{1}{2}} = (\Omega_{A_1} + \gamma_r) p^e_{\pm \frac{1}{2}} + \gamma_3 p_{ms_1} + \gamma'_{3,0} - \Omega_{A_1} p^g_{\pm \frac{1}{2}} \qquad (SI\ 27)$$

Here $p_i^{e(g)}$ are the excited (ground) state population for $i = \{\pm\frac{1}{2}, \pm\frac{3}{2}\}$. Inherent, first-order power-independent, decay from MS$_2$ to the ground states is taken into account via $\gamma'_{3,0}$ and $\gamma'_{4,0}$. $R$ describes the rate from MS$_2$ to MS$_3$, where the explicit power-dependence will be assigned later. Under the assumption that decay from MS$_3$ to the excited states happens fast, i.e. $R \ll \Gamma$, the population of MS$_3$ can be adiabatically eliminated, yielding

$$p_{ms_3} = \frac{R}{2\Gamma}. \qquad (SI\ 28)$$

As the Rabi frequencies corresponding to the applied laser powers are much lower than $\Gamma_1, \Gamma_2$, also the excited state population can be eliminated adiabatically,

$$p^e_{\pm 1/2} = \frac{1}{2}\frac{R}{\Gamma_1 + \Omega_{A_1}} p_{ms_2} + \frac{1}{2}\frac{\Omega_{A_1}}{\Gamma_1 + \Omega_{A_1}} p^g_{\pm 1/2}$$

$$p^e_{\pm \frac{3}{2}} = \frac{1}{2}\frac{R}{\Gamma_2 + \Omega_{A_2}} p_{ms_2} + \frac{1}{2}\frac{\Omega_{A_2}}{\Gamma_2 + \Omega_{A_2}} p^g_{\pm \frac{3}{2}} \qquad (SI\ 29)$$

With this, the initial set of rate equations can be reduced to

$$\dot{p}_{ms_2} = -(\gamma'_3(R) + \gamma'_4(R) + \kappa)p_{ms_2} + P'_{A_1}p^g_{\pm\frac{1}{2}} + P'_{A_2}p^g_{\pm\frac{3}{2}}$$

$$\dot{p}_{ms_1} = \kappa p_{ms_2} + P_{A_1}p^g_{\pm 1/2} + P_{A_2}p^g_{\pm 3/2} - (\gamma_3 + \gamma_4)p_{ms_1}$$

$$\dot{p}^g_{\pm 3/2} = \gamma'_4(R)p_{ms_2} - (P_{A_2} + P'_{A_2})p^g_{\pm\frac{3}{2}} + \gamma_4 p_{ms_1}$$

$$\dot{p}^g_{\pm\frac{1}{2}} = \gamma'_3(R)p_{ms_2} - (P_{A_1} + P'_{A_1})p^g_{\pm\frac{1}{2}} + \gamma_3 p_{ms_1} \qquad (SI\ 30)$$

Here, the following parameters have been defined:

$$P^{(\prime)}_{A_1} \equiv \frac{\Omega_{A_1}\gamma^{(\prime)}_1}{\Omega_{A_1} + \Gamma_1} \qquad P^{(\prime)}_{A_2} \equiv \frac{\Omega_{A_2}\gamma^{(\prime)}_2}{\Omega_{A_2} + \Gamma_2}$$

$$\kappa \equiv \frac{R}{2}\left(\frac{\gamma_1}{\Omega_{A_1} + \Gamma_1} + \frac{\gamma_2}{\Omega_{A_2} + \Gamma_2}\right)$$

$$\gamma'_3(R) = \gamma'_{3,0} + \frac{R}{2}\frac{\Omega_{A_1} + \gamma_r}{\Omega_{A_1} + \Gamma_1}$$

$$\gamma'_4(R) = \gamma'_{4,0} + \frac{R}{2}\frac{\Omega_{A_2} + \gamma_r}{\Omega_{A_2} + \Gamma_2} \qquad (SI\ 31)$$

If one assumes deshelving from MS$_2$ to MS$_3$ to be direct proportional to the drives along A$_1$ and A$_2$, then one can define

$$R = \beta(\Omega_{A_1} + \Omega_{A_2}) \qquad (SI\ 32)$$

with $\beta$ describing the efficiency of the excitation. With this, one ends up with

$$\gamma'_3(\Omega_{A_1}, \Omega_{A_2}) = \gamma_{3,0} + \beta\frac{\Omega_{A_1} + \Omega_{A_2}}{2}\frac{\Omega_{A_1} + \gamma_r}{\Omega_{A_1} + \Gamma_1}$$

$$\gamma'_4(\Omega_{A_1}, \Omega_{A_2}) = \gamma'_{4,0} + \beta\frac{\Omega_{A_1} + \Omega_{A_2}}{2}\frac{\Omega_{A_2} + \gamma_r}{\Omega_{A_2} + \Gamma_2} \qquad (SI\ 33)$$

### SI10. Saturation Curve

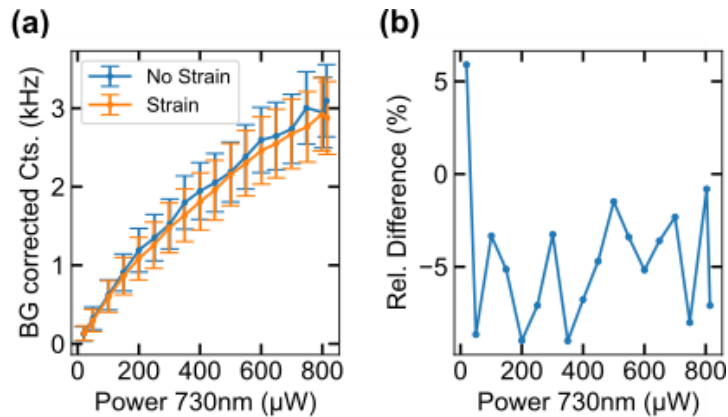

*Figure SI 9: (a) Background-corrected saturation curve of two $V_{Si}$ centers with and without strain. (b) Relative difference in the photoluminescence from (a) of the strained $V_{Si}$ with respect to the unstrained $V_{Si}$ indicating a strain-induced reduction of the photoluminescence by ≈-5%.*

To support the finding of reduced photon emission due to strain we measured a saturation curve for two emitters with and without strain. Hereby the off-resonant laser power at 730 nm was increased up to 815 µW. The measured background-corrected photoluminescence can be seen in Fig. (SI) 9(a). In accordance with the simulations presented in the main text, we find a strain-induced reduction of the emitted photoluminescence by $\approx -5\%$.